\documentclass[a4paper, 10 pt]{amsart}  
\pdfoutput=1 
\usepackage{graphicx}    
\usepackage{times} % assumes new font selection scheme installed
\usepackage{amsmath} % assumes amsmath package installed
\usepackage{amssymb}  
\usepackage{cite}
\usepackage{tikzit}
\usepackage[hidelinks]{hyperref}
\usepackage{easyReview}
\usepackage{float}
\usepackage{amsthm}

\usepackage{subcaption}

\tikzstyle{Box}=[fill=white, draw=black, shape=rectangle, minimum height=10 mm, thick, minimum width=10 mm]
\tikzstyle{new style 0}=[fill=white, draw=black, shape=circle, minimum height=2mm, thick]

\tikzstyle{arrow}=[->, thick, >=stealth]
\tikzstyle{line}=[-, thick]
\tikzstyle{new edge style 0}=[-, thick, dashed]
\tikzstyle{Box_edge}=[-, thick]

\usepackage{amsthm}
\usepackage{mathrsfs}
\usepackage{empheq}
\usepackage{tikz}

\newtheorem{theorem}{Theorem}

\newtheorem{lemma}{Lemma}
\newtheorem{definition}{Definition}

\usepackage{eso-pic}
\AddToShipoutPictureBG*{%
\AtPageUpperLeft{%
\setlength\unitlength{1in}%

\hspace*{\dimexpr0.5\paperwidth\relax}
\makebox(0,-1.75)[c]{
\begin{tabular}{c c}
Timo de Groot \emph{et al.},
EXPLOITING STRUCTURE IN MIMO SCALED
GRAPH ANALYSIS, \\
uploaded to ArXiv  April 14, 2025\\
\end{tabular}}}}

\usepackage[foot]{amsaddr}

\title{\LARGE \bf Exploiting Structure in MIMO Scaled Graph Analysis}
\author{Timo de Groot$^1$, Tom Oomen$^{1,2}$, Sebastiaan van den Eijnden$^1$}
\address{$^1$ All authors are with the Department of Mechanical Engineering, Eindhoven University of Technology,  5600 MB Eindhoven, The Netherlands. E-mail: $\{$t.d.groot2, t.a.e.oomen, s.j.a.m.v.d.eijnden$\}$@tue.nl.}
\address{$^2$ T. Oomen is also with the Department of Delft Center for Systems and Control, Delft University of Technology, Delft, The Netherlands}

\date{}

\begin{document}

\maketitle
\thispagestyle{empty}
\pagestyle{empty}

%%%%%%%%%%%%%%%%%%%%%%%%%%%%%%%%%%%%%%%%%%%%%%%%%%%%%%%%%%%%%%%%%%%%%%%%%%%%%%%%

\textbf{\textit{Abstract} --- Scaled graphs offer a graphical tool for analysis of nonlinear feedback systems. Although recently substantial progress has been made in scaled graph analysis, at present their use in multivariable feedback systems is limited by conservatism. In this paper, we aim to reduce this conservatism by introducing multipliers and exploit system structure in the analysis with scaled graphs. In particular, we use weighted inner products to arrive at a weighted scaled graph and combine this with a commutation property to formulate a stability result for multivariable feedback systems. We present a method for computing the weighted scaled graph of Lur'e systems based on solving sets of linear matrix inequalities, and demonstrate a significant reduction in conservatism through an example.  }

\section{Introduction}
Many feedback control systems are multivariable and nonlinear, requiring specialized analysis and design tools. For linear time-invariant (LTI) feedback systems, tools like Nyquist and Bode plots provide effective graphical methods for transparent and intuitive analysis and design \cite{skogestad_multivariable_2005}. Recently, the concept of the Nyquist plot has been extended to the nonlinear domain through the introduction of the scaled graph \cite{ryu_scaled_2022, chaffey_graphical_2023, chaffey_loop_2022}. Scaled graphs originate from the optimization literature and have been developed to provide rigorous and intuitive proofs for the convergence of many convex optimization algorithms \cite{ryu_scaled_2022}. Since their introduction to the control community, scaled graphs have been applied in various systems and control contexts, including Lur’e systems \cite{krebbekx_srg_2024}, nonmonotone operators \cite{quan_scaled_2024}, multivariable LTI control systems \cite{baron-prada_mixed_2025, baron-prada_stability_2025}, and reset control systems \cite{van_den_eijnden_scaled_2024}.

Although substantial progress has been made in scaled graph analysis, and it is recently broadly adopted for control design, at present the use of scaled graphs for multivariable (nonlinear) systems is hampered by conservatism. This conservatism seems attributed to the fact that in their current definition, scaled graphs do not allow for directly incorporating structural knowledge of the system. Exploiting structural knowledge of components in a feedback loop is well-known to reduce conservatism within the context of, e.g., robustness analysis of multivariable LTI systems \cite{packard_complex_1993, skogestad_multivariable_2005, doyle_analysis_1982}. 

The aim of this paper is to pinpoint the conservatism in multivariable scaled graph analysis, and effectively reduce it via a new approach that makes use of scaling in the feedback loop. Our approach is akin to classical multiplier techniques \cite{scherer_robust_2007, turner_analysis_2020}, and makes use of weighted inner products. The structure of the weights relates to specific structural properties of the feedback components (e.g., a diagonal structure), allowing for structural information to be embedded within the weighted scaled graph.

In line with the above, the main contributions of this paper are as follows. First, we define the weighted scaled graph and formulate a stability result for multivariable feedback systems in which we exploit a commutation property of one of the feedback components. Second, we present a method for constructing (an over-approximation of) the weighted scaled graph for the specific class of Lur'e control systems \cite{khalil_nonlinear_2014}. Lur'e control systems form an important class of systems that arise in many applications, ranging from neural networks to nonlinear motion control systems \cite{van_den_eijnden_frequency-domain_2023, heertjes_self-tuning_2013,yin_design_2011,pauli_linear_2021}. Our approach for computing the weighted scaled graphs of Lur'e systems is inspired by \cite{van_den_eijnden_scaled_2024}, and is based on solving specific sets of linear matrix inequalities (LMIs). Feasibility of these LMIs guarantees that the system under consideration satisfies a number of specific integral quadratic constraints (IQCs) \cite{megretski_system_1997}. Each IQC translates to a region in the complex plane that over-approximates the weighted scaled graph. By taking the intersection of the regions described by the different IQCs, we further reduce the over-approximation, leading to a tighter estimation of the weighted scaled graph. We demonstrate effectiveness of the computational method as well as the possibility to significantly reduce conservatism through weighted scaled graphs in an example. 

The remainder of this paper is organized as follows. In Section~\ref{sec:prelim}, preliminaries are given and the problem statement is discussed. In Section~\ref{sec:main1}, we present our first main contribution in the form of stability conditions for multivariable feedback systems based on weighted scaled graphs. Section~\ref{eq:comp} presents our second main contribution which entails a computational method for obtaining the weighted scaled graph of Lur'e systems by solving sets of LMIs. We demonstrate the effectiveness of our approaches through an example in Section~\ref{sec:lure_example}. The main conclusions are summarized in Section~\ref{sec:conclusions}.

\textbf{Notation.} The sets of $n$-by-$n$ real symmetric matrices are denoted by $\mathbb{S}^{n}=\{W\in\mathbb{R}^{n\times n}\mid W=W^\top\}$. For $W \in \mathbb{S}^n$, we use $W\succ 0$ (resp. $W\succeq 0$) to indicate that $W$ is positive definite (resp. positive semi-definite), i.e., $x^\top W x>0$ for all $x\in\mathbb{R}^{n}\setminus\{0\}$ (resp. $x^\top W x \geq 0$ for all $x\in \mathbb{R}^n$). We will use a similar convention for negative (semi-)definite matrices. Furthermore, $\otimes$ denotes the Kronecker product and $\text{diag}(a_1,...,a_n)$ denotes a matrix with $a_1,...,a_n$ on the diagonal and $0$ elsewhere.

For signals $u,y :[0,\infty) \to \mathbb{R}^n$ and a positive (semi-) definite matrix $W \in \mathbb{S}^n$ we denote 
\begin{equation*}
    \langle u,y \rangle_W = \int_{0}^{\infty}\!\!\!\!\! u(t)^\top W y(t)\; dt, \:\: \textup{ and } \:\: \|u\|_W^2 = {\langle u,u\rangle_W}.
\end{equation*}
When $W = I$, we write $\langle u,y\rangle_W = \langle u,y\rangle$ and $\|u\|_W = \|u\|$. The space of signals that are square-integrable on the complete time axis $[0,\infty)$, i.e., signals that satisfy $\|u\|< \infty$ is denoted by $\mathcal{L}_2^n$. The space of signals that are square-integrable on any finite time interval $[0,T]$, i.e, signals that satisfy $\int_{0}^T u^\top (t)u(t)dt < \infty$ for all $T\geq 0$  is denoted by $\mathcal{L}_{2e}^n$. We denote the distance between two sets $A, B \subset \mathbb{C}$ by $\textup{dist}(A,B) = \inf_{a \in A, b \in B}|a-b|$. For a complex number $z=a+bj$, we denote its complex conjugate by $z^*=a-bj$.

\section{Preliminaries and Problem Statement}\label{sec:prelim}
\subsection{System Setting}
Consider a nonlinear system $H$ represented in state-space form by
\begin{align} \label{eq:canonical_nl_system}
H:\begin{cases}
    \dot{x} = f(x,u), \qquad x(0) = 0,\\
    y = g(x,u),
\end{cases}
\end{align} 
with state $x\in\mathbb{R}^m$, input $u\in\mathbb{R}^n$, output $y\in\mathbb{R}^n$, and nonlinear functions $f:\mathbb{R}^m\times\mathbb{R}^n\rightarrow\mathbb{R}^m$ and $g:\mathbb{R}^m \times\mathbb{R}^n\rightarrow\mathbb{R}^n$ which satisfy $f(0, 0)=0$, and $g(0, 0)=0$. System \eqref{eq:canonical_nl_system} is square, as it has an equal number of inputs and outputs. Solutions to \eqref{eq:canonical_nl_system} are considered as
absolutely continuous functions $x:[0,T]\to \mathbb{R}^m$ that satisfy
\eqref{eq:canonical_nl_system} for almost all times $t\in[0,T]$. We assume that $f$ satisfies certain regularity properties such that global
existence of solutions is guaranteed \cite{khalil_nonlinear_2014}. We write $y \in H(u)$ to denote (possibly multi-valued) outputs of \eqref{eq:canonical_nl_system} resulting from inputs $u$, and we write $\tau H$ to indicate scaling of $y$ by a number $\tau$.

We will consider stability properties of (feedback interconnections of) systems as in \eqref{eq:canonical_nl_system}, which we formalize next.

\begin{definition}
    A system of the form \eqref{eq:canonical_nl_system} is said to be stable if inputs $u \in \mathcal{L}_2^n$ are mapped to outputs $y \in \mathcal{L}_2^n$. It is said to be finite-gain stable if it is stable and there exists $\gamma >0$ such that $\|y\|\leq \gamma \|u\|$. Here, $\gamma$ is denoted the $\mathcal{L}_2$-gain. \hfill $\ulcorner$
\end{definition}

\subsection{Scaled Graphs}
Let $u \in \mathcal{L}_2^n$ and suppose that $H$ in \eqref{eq:canonical_nl_system} is stable, i.e., $H:\mathcal{L}_2^n \rightarrow \mathcal{L}_2^n$. The scaled graph of $H$ is defined as \cite{chaffey_graphical_2023}
\begin{equation}\label{eq:SG}
    \textup{SG}(H) = \left\{\rho(u,y)e^{\pm j\theta(u,y)} \mid u \in \mathcal{L}_2, \:y \in H(u)\right\},
\end{equation}
with gain $\rho(u,y) = \|y\|/\|u\|$ if $u,y\neq 0$ and $\rho(u,y) = \infty$ when $u = 0$, and phase $\theta(u,y) = \arccos\left(\tfrac{\langle u,y\rangle}{\|u\|\|y\|}\right)$ if $u,y\neq 0$ and $\theta(u,y) = 0$ if $u=0$. The scaled graph in \eqref{eq:SG} generalizes the idea of a Nyquist plot by representing input-output information of a system through complex numbers having gain and phase. The inverse of the SG in \eqref{eq:SG} is denoted by $\textup{SG}^\dagger (H)$ and is obtained by swapping the role of the input and output of $H$, leading to
\begin{align} 
    \textup{SG}^\dagger(H) = \left\{\frac{1}{\rho(y,u)}e^{\pm j\theta(y,u)} \mid u \in \mathcal{L}_2, \:y \in H(u)\right\}.
\end{align} 

The next result from \cite{van_den_eijnden_scaled_2024} characterizes stability of the negative feedback interconnection of two systems $H_1$ and $H_2$ in terms of the separation of their scaled graphs. 

\begin{theorem}\label{th:SG}
Consider a pair of finite-gain stable systems $H_1$ and $H_2$, and suppose that the negative feedback interconnection of $H_1$ and $\tau H_2$ is well-posed\footnote{A feedback interconnection is well-posed if, given input signals in $\mathcal{L}_2$, there exist output signals in $\mathcal{L}_{2e}$ depending causally on the inputs, see \cite{freeman_role_2022}.} for all $\tau \in (0,1]$. If there exists $r >0$ such that for all $\tau \in (0,1]$ 
\begin{equation}\label{eq:cond}
   \textup{dist}(\textup{SG}^\dagger(H_1), \textup{SG}(-\tau H_2))\geq  r
\end{equation}
then the feedback interconnection is finite-gain stable with an $\mathcal{L}_2$ gain bound of $1/r$. \hfill $\ulcorner$
\end{theorem}

Condition \eqref{eq:cond} can be checked graphically, which makes Theorem~\ref{th:SG} interesting also from a practical perspective. For single-input single-output (SISO) systems, the check is akin to a classical Nyquist stability test. While Theorem~\ref{th:SG} also applies to multi-input multi-output (MIMO) systems, we argue that this result is inherently conservative due to the definition of the scaled graph in \eqref{eq:SG}. This conservatism is revealed through a simple example in the next section.

\subsection{Example and Problem Statement}
We illustrate conservatism in scaled graph analysis for MIMO systems through an LTI example. Although The LTI case is a special case of the nonlinear case, it allows for comparison with exact stability tests, such as the generalized Nyquist criterion.

\begin{figure}[t]
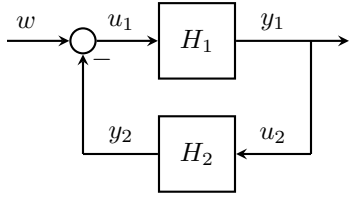

	\centering
	\ctikzfig{General_FB_interconnection}
	\caption{General feedback interconnection.}
	\label{fig:General_FB}
\end{figure}

Consider the feedback interconnection shown in Figure~\ref{fig:General_FB}, where $H_1$ and $H_2$ are represented by the transfer function matrices
\begin{align*} 
H_1(s) = \begin{bmatrix} \frac{5}{s^2+2s+1}&\!\!\!\!\! \frac{10}{s+1} \\ 0& \!\!\!\!\! \frac{1}{s+1}\end{bmatrix},  \textup{ and } H_2(s) = \begin{bmatrix} \frac{1}{s+1}&\!\!\!\!\!0\\0& \!\!\!\!\!\frac{5}{s^2+4s+4} \end{bmatrix}.  
\end{align*}

Using well-known methods from linear system theory \cite{skogestad_multivariable_2005}, it is straightforward to verify that the negative feedback interconnection in Figure~\ref{fig:General_FB} between $H_1$ and $H_2$ is stable. Let $y_1^\top = [y_{11}\,\, y_{12}]$ and note that $y_1 =H_1(u_1)$. Take the specific input $u_1(t)^\top  = \begin{bmatrix} u_{11}(t)&\!\!\!\!0\end{bmatrix}  $ with $u_{11} \in \mathcal{L}_2$ having its Fourier coefficients centered on frequency $\pm\omega_0$ such that
\begin{equation}
    \hat{u}_{11}(j\omega) = \begin{cases}
        c & \textup{if } |\omega - \omega_0|<\epsilon \textup{ or } |\omega + \omega_0|<\epsilon \\
        0 & \textup{otherwise},
    \end{cases}
\end{equation}
where $c$ is chosen such that $\|u_1\|=\|u_{11}\| = 1$, and $\epsilon>0$ is sufficiently small. Then,
\begin{equation}
    \rho_{H_1}(u_1,y_1) := \frac{\|y_1\|}{\|u_1\|} = \|y_1\| = |H_{1,11}(j\omega_0)|,
\end{equation}

and
\begin{equation}
    \theta_{H_1}(u_1,y_1) := \arccos\left(\frac{\langle u_1,y_1\rangle}{\|u_{1}\|\|y_{1}\|}\right) = \angle(H_{1,11}(j\omega_0)),
\end{equation}
such that the complex numbers
\begin{equation}
   z_{H_1} = \frac{1}{\rho_{H_1}(u_1,y_1)}e^{ \pm j\theta_{h_1}(u_1,y_2)}\in \textup{SG}^\dagger(H_1).
\end{equation}
Furthermore, using a similar approach it is easy to verify that the complex numbers
\begin{equation}
   z_{H_2} = -|H_{2,22}(j\omega_1)|e^{\pm j \angle(H_{2,22}(j\omega_1))} \in \textup{SG}(-H_2).
\end{equation}
\noindent Taking, for example, $\omega_0 = 1$ rad/s and $\omega_1 = 2$ rad/s results in phases $|\angle(z_{H_1})|=|\angle(z_{H_2})|$, and gains $|z_{H_1}| = 0.4$, and $|z_{H_2}| = 0.625$. This shows that there exists $\tau \in (0,1]$ such that $|z_{H_1} -\tau z_{H_2}| = 0$ ($\tau=0.64$ for this example). In other words, $\textup{SG}^\dagger(H_1)$ and $\textup{SG}(-\tau H_2)$ overlap, and condition \eqref{eq:cond} in Theorem~\ref{th:SG} is not satisfied. The reason for the inherent conservatism in this example is two-fold. First, whereas in a Nyquist analysis only interaction at corresponding frequencies is considered, i.e., it is sufficient to look at $\omega_0 = \omega_1$ in the example, SG analysis handles interactions across different frequencies, i.e., it considers a worst-case situation where all possible input-output pairs are matched. Second, the diagonal structure of $H_2$ is not explicitly taken into account in the analysis, again leading to a worst case analysis. If the structure of $H_2$ is explicitly taken into account for this example, stability can be verified using standard methods from, e.g., \cite{skogestad_multivariable_2005}.

While the first source of conservatism can be reduced for LTI systems by considering recently introduced frequency-wise SGs \cite{baron-prada_mixed_2025,baron-prada_stability_2025}, this is not trivial for nonlinear systems. The second source of conservatism, however, can be significantly alleviated in both the LTI and nonlinear case through appropriate input-output scaling, as we show in this paper. Our idea is largely inspired by the use of multipliers (so-called D-scalings) in analysis of MIMO LTI feedback systems subject to structured uncertainties \cite{packard_complex_1993}.

\section{Weighting in Scaled Graphs}\label{sec:main1}
In this section we introduce multipliers in scaled graphs, and present our first main result in Theorem \ref{th:SGW}, in the form of conditions for feedback stability.

\subsection{Weighted Inner Products in Scaled Graphs}
Consider a system $y = H(u)$, with $u,y \in \mathcal{L}_2^n$. For some nonsingular matrix $X \in \mathbb{R}^{n\times n}$ we denote the input-output transformation $\bar{u} = Xu$ and $\bar{y} = Xy$, leading to the transformed system $$\bar{y} = \bar{H}(\bar{u}) = XH(X^{-1}\bar{u}),$$ as shown in Figure~\ref{fig:first_loop_transform}.   
For simplicity, we write $\bar{H} = XHX^{-1}$. From the fact that $\langle \bar{u}, \bar{y} \rangle = \langle u,y\rangle_W$, $\|\bar{y}\| = \|y\|_W$, and $\|\bar{u}\|=\|u\|_W$ in which $W=X^\top X$, it immediately follows that the scaled graph of $\bar{H}$ can be expressed in terms of the weighted inner products of the original signals $u$ and $y$. This leads to the following weighted version of the scaled graph of $H$ in \eqref{eq:SG} as
\begin{equation}\label{eq:SGW}
    \textup{SG}^W\!(H) \!= \!\! \left\{ \!\rho_W(u,y)e^{\pm j\theta_W(u,y)} \!\! \mid \! u \! \in \! \mathcal{L}_2, y \! \in \! H(u)\right\},
\end{equation}
with gain $\rho_W(u,y) = \|y\|_W/\|u\|_W$ and phase $\theta_W(u,y) = \arccos\left(\tfrac{\langle u,y\rangle_W}{\|u\|_W\|y\|_W}\right)$. 

Clearly, in the case where $H$ is a SISO system, or when $H$ is a MIMO system and $X=I$, we recover $\textup{SG}(H)$. In addition, for certain special classes of MIMO systems we may also recover $\textup{SG}(H)$ for $X\neq I$. This is in particular the case when $H$ and $X$ commute, which we define as follows.

\begin{definition}
    A system $H: \mathcal{L}_2^n \to \mathcal{L}_2^n$ commutes with a matrix $X$ if 
    \begin{equation}
        H(Xu) = XH(u).
    \end{equation}
\end{definition}
We denote the set of all systems that commute with matrices $X \in \mathcal{X}$ by $\mathcal{C}(\mathcal{X})$. Note that a commutativity property implies that $H$ has a certain structure, e.g., $H$ may be a diagonal system. We have the following result.
\begin{lemma}\label{eq:lem1}
    Let $H \in \mathcal{C}(\mathcal{X})$. Then for $X \in \mathcal{X}$ we have $\textup{SG}^W(H) = \textup{SG}(H)$.\hfill $\ulcorner$
\end{lemma}
All $n\times n$ diagonal LTI systems commute with diagonal matrices $X =\textup{diag}(x_1, \ldots, x_n)$. Moreover, any diagonal nonlinear system that is homogeneous of degree one commutes with $X =\textup{diag}(x_1, \ldots, x_n)$. Relevant examples of such nonlinearities are given by reset and hybrid integrators, see, e.g., \cite{van_loon_split-path_2016, van_den_eijnden_hybrid_2020, foster_nonlinear_1966}. 

\subsection{Feedback Stability}
\begin{figure}[t]
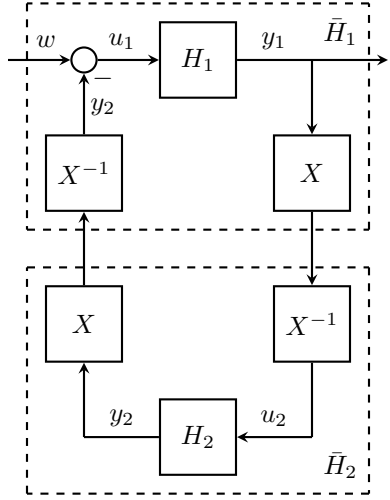

	\centering
	\ctikzfig{First_step_Loop_transform}
	\caption{Transformed feedback interconnection.}
	\label{fig:first_loop_transform}
\end{figure}

Equipped with the above results, we are now ready to formulate a feedback stability theorem for the general feedback interconnection depicted in Figure~\ref{fig:General_FB}.
\begin{theorem}\label{th:SGW}
Consider a pair of finite-gain stable systems $H_1$ and $H_2$, and suppose that the negative feedback interconnection of $H_1$ and $\tau H_2$ is well-posed for all $\tau \in (0,1]$. Furthermore, suppose that for all $\tau \in (0,1]$ $\tau H_2 \in \mathcal{C}(\mathcal{X})$. If there exists $r >0$ such that for all $\tau \in (0,1]$ and some $W=X^\top X$ with $X \in \mathcal{X}$
\begin{equation}\label{eq:cond_Ww}
   \textup{dist}(\textup{SG}^{W\dagger}(H_1), \textup{SG}(-\tau H_2))\geq  r
\end{equation}
then the feedback interconnection is finite-gain stable with an $\mathcal{L}_2$ gain bound of $\sqrt{\bar{\lambda}(W)}/$\\$(r\sqrt{\underline{\lambda}(W)})$, where $\bar{\lambda}(W)$ and $\underline{\lambda}(W)$ denote the maximum and minimum eigenvalue of the matrix $W$ respectively. \hfill $\ulcorner$
\end{theorem}
\begin{proof}
    Since $\tau H_2 \in \mathcal{C}(\mathcal{X})$ we find from Lemma~\ref{eq:lem1} that $\textup{SG}(-\tau H_2) = \textup{SG}^W(-\tau H_2)$. Hence, condition \eqref{eq:cond_Ww} implies via Theorem~\ref{th:SG} that the feedback interconnection of $XH_1X^{-1}$ and $XH_2X^{-1}$ is finite-gain stable. Since this is equivalent to the interconnection of $H_1$ and $H_2$ (see Figure~\ref{fig:first_loop_transform}),  we arrive at the stability result. Furthermore, from Theorem~\ref{th:SG} we get $\|\bar{y}_1\|\leq \tfrac{1}{r} \|\bar{w}\|$. Since $W\succ0$, $\|{y_1}\|_W=\|\bar{y}_1\|$ and $\|\bar{w}\|=\|w\|_W$ we get $\sqrt{\underline{\lambda}(W)} \|y_1\|\leq \|\bar{y_1}\|\leq \tfrac{1}{r} \|\bar{w}\|\leq \tfrac{\sqrt{\bar{\lambda}(W)}}{r} \|w\|$, with $\underline{\lambda}(W)$ and $\bar{\lambda}(W)$ the smallest and largest eigenvalue of $W$, respectively. Thus, we arrive at the bound on the $\mathcal{L}_2$ gain.  \hfill $\square$
\end{proof}

\begin{figure}[b]
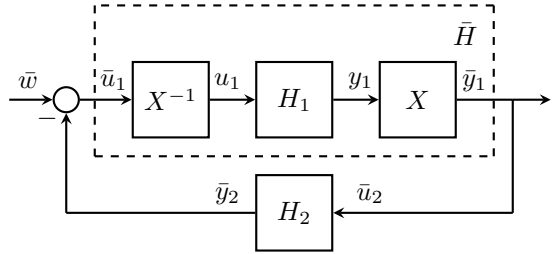

	\centering
	\ctikzfig{Final_loop_transformation}
	\caption{Transformed feedback interconnection with $H_2\in\mathcal{C}(\mathcal{X})$.}
	\label{fig:final_loop_transform}
\end{figure}

Theorem~\ref{th:SGW} essentially exploits the loop-transformations depicted in Figure~\ref{fig:first_loop_transform} and Figure~\ref{fig:final_loop_transform}. This approach is closely related to multiplier techniques in nonlinear system analysis \cite{turner_zames-falb_2022, turner_analysis_2020, desoer2009feedback}. The use of (frequency-wise) multipliers and D-scalings, and exploitation of commutation properties relates to classical robust stability and performance analysis of MIMO LTI systems with structured uncertainty \cite{packard_complex_1993, poola_robust_1995, scherer_robust_2007}. Although in this paper we focus on static multipliers with the aim of exploiting structure in MIMO systems, the results can be lifted to the use of dynamic multipliers as well. The benefits of static matrix multipliers comes, amongst others, from the possibility to find these multipliers using efficient numerical procedures, as we show in the next section for the specific class of Lur'e systems.  

\section{Computations for Lur'e Systems}\label{eq:comp}
In this section, we explore the previous results within the context of MIMO Lur'e systems, and focus in particular on computation of the weighted scaled graph.

\subsection{Lur'e System Description}
\begin{figure}[t]
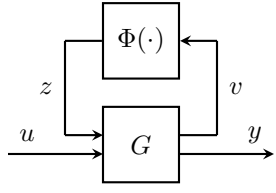

	\centering
	\ctikzfig{Lure_system_description}
	\caption{MIMO Lur'e system, where $G$ represents the LTI part of the system and $\Phi(\cdot)$ is a MIMO nonlinearity that satisfies the IQC in \eqref{eq:lure_IQC}.}
	\label{fig:final_NL_loop_transform}
\end{figure}

Consider a MIMO Lur'e system $H$ described by
\begin{align} \label{eq:canonical_lure}
H:\left\{\begin{aligned}
    \dot{x} &= Ax+B_z z+ B_u u, \quad\quad\quad x(0) = 0,\\
    v &= C_vx + D_{zv}z + D_{uv} u,\\
    y &= C_yx+D_{zy}z + D_{uy} u,\\
    z& = \Phi(v),
\end{aligned}\right.
\end{align} 
where $x\in\mathbb{R}^n$ denotes the state, $u\in \mathbb{R}^q, z\in\mathbb{R}^p$ denote the inputs and $y\in \mathbb{R}^q, v\in\mathbb{R}^{p}$ denote the outputs of the LTI part of the system. The nonlinearity $\Phi:\mathcal{L}_{2}^p \to \mathcal{L}_2^p$ is assumed to be a (possibly dynamical) system that satisfies the integral quadratic constraint (IQC) \cite{megretski_system_1997} 
\begin{align} \label{eq:lure_IQC}
\int_0^{\infty}\begin{bmatrix} z(t)\\v(t) \end{bmatrix}^\top \Theta \otimes I_p\begin{bmatrix} z(t) \\ v(t) \end{bmatrix} \;dt \geq 0,  
\end{align} 
for some given matrix $\Theta \in \mathbb{S}^{2}$. The Lur'e system is schematically depicted in the block diagram in Figure~\ref{fig:final_NL_loop_transform}.  

\subsection{From LMIs to Scaled Graphs}
Our next result links an over-approximation of the weighted scaled graph of the Lur'e system $H$ in \eqref{eq:canonical_lure} to an LMI condition.

\begin{theorem}\label{th:Lure_LMI}
Consider the system $H$ in \eqref{eq:canonical_lure} and suppose that this system is stable. Suppose that for given matrices $\Theta, \Pi \in \mathbb{S}^2$ there exist matrices $W, P \in \mathbb{S}^q$ and a number $\sigma \geq 0$ that satisfy the LMI conditions
\begin{subequations}\label{eq:Lure_LMI}
  \small  \begin{align}
        W& \succ 0,  \\
  \Gamma_P^\top \begin{bmatrix} 0 & P\\P & 0 \end{bmatrix}\Gamma_P +\Gamma_y^\top\Pi\otimes W\Gamma_y +\sigma\Gamma_v\Theta\otimes I_p \Gamma_v & \preceq 0,     
    \end{align}
\end{subequations}
with
\begin{align} \label{eq:Gamma_matrices}
\Gamma_P^\top =\!\begin{bmatrix} A^\top &\!\!\! I \\B_z^\top &\!\!\! 0 \\B_u^\top &\!\!\! 0 \end{bmatrix}\!, \; \Gamma_y^\top = \!\begin{bmatrix} C_y^\top& \!\!\!0 \\ D_{zy}^\top &\!\!\! 0 \\ D_{uy}^\top &\!\!\! I \end{bmatrix}\! , \; \Gamma_v^\top = \!\begin{bmatrix} 0 &\!\!\! C_v^\top \\ I &\!\!\! D_{zv}^\top\\0&\!\!\!D_{uv}^\top\end{bmatrix}.
\end{align} 
Then, $\textup{SG}^{W}(H) \subseteq \mathcal{S}(\Pi)$, in which 
\begin{align} \label{eq:SP}
\mathcal{S}(\Pi)=\left\{z\in\mathbb{C}\;\left| \begin{bmatrix} z \\ 1 \end{bmatrix}^* \Pi \begin{bmatrix} z \\ 1 \end{bmatrix} \leq 0 \right.\right\}.
\end{align} 
\end{theorem}

 The proof is postponed to Appendix~\ref{app:A}.

Theorem~\ref{th:Lure_LMI} provides a computationally efficient method for constructing (an over-approx-imation of) $\textup{SG}^W(H)$ based on a given $\Pi$. Of particular interest are matrices of the form
\begin{equation}\label{eq:PI}
\Pi = \pm \begin{bmatrix} 1 & -\lambda \\-\lambda & \lambda^2-\rho^2 \end{bmatrix},
\end{equation}
with $\lambda, \rho \in \mathbb{R}$. The matrix $\Pi$ in \eqref{eq:PI} describes the interior of a circle in $\mathbb{C}$ with center point $\lambda$ and radius $\rho$, and $-\Pi$ describes the exterior of such circle. By solving $N$ LMIs in Theorem~\ref{th:Lure_LMI} for different matrices $\Pi=\Pi_i$, $i = \left\{1,\ldots N\right\}$ (while keeping $W$ fixed for each $i$), we obtain an over approximation of $\textup{SG}^W(H)$ by taking the intersection of the different regions $\mathcal{S}(\Pi_i)$ as in \eqref{eq:SP}. 

\subsection{Finding a Suitable $W$}
%\textcolor{red}{Dit moet wat duidelijker. Hoe zoeken we Pi? W is gefixt over alle Pi.}
In this section, a method for finding suitable values for $W$ is discussed.
Since $\Pi$ and $W$ appear as products in \eqref{eq:Lure_LMI}, both $\Pi$ and $W$ cannot be free variables simultaneously. Therefore, a bisection search \cite{fepperson_introduction_2013} is used to find suitable values for $W$. We keep $W$ to have a diagonal structure, and fix $\Pi = \textup{diag}(1,-\gamma)$. We initially set $W = I$, and solve the LMIs with $\gamma$ a free variable, the latter characterizing an upper-bound on the $\mathcal{L}_2$-gain of the system. Next, a bisection is performed over $\gamma$, while iteratively solving \eqref{eq:Lure_LMI} with $W$ a free variable (but restricted to be a diagonal matrix). The final $W$ resulting in the smallest $\gamma$ is then fixed in the LMIs, and we proceed to solve these LMIs for $\Pi_i$, $i = \left\{1,\ldots, N\right\}$. 
The $\Pi_i$ matrices are chosen as in \eqref{eq:PI} with different values for the corresponding $\lambda_i$. When solving each LMI, $\rho_i^2$ is maximized (when $\Pi=\Pi_i$) and minimized (when $\Pi = -\Pi_i$) to find in each $i$-th iteration the smallest circle centered at $\lambda_i$ that must contain the scaled graph, and, respectively, find the largest circle that does not contain the scaled graph.

\subsection{Discussion}
The graphical nature of scaled graphs provide several unique benefits over direct use of LMIs both for SISO and MIMO systems. 1) In robustness analysis, classes of allowable uncertainties (expressed in terms of properties of their scaled graphs) and robustness measures can be directly read off the scaled graph. We highlight this in an example below. In contrast, such information is not easily extracted from the LMIs. 2) Regarding controller design, LMIs offer limited guidance into the (re)design and tuning of robustly stabilizing controllers. On the other hand, scaled graphs offer direct insights into classes of stabilizing controllers (in terms of properties of their scaled graphs). 3) Finally, the above construction of scaled graphs is based on taking the intersection of several IQCs, which can be interpreted as taking non-convex combinations of IQCs. Classical methods, on the other hand, rely on taking convex combinations (or in other words: taking the union) of several IQCs \cite{megretski_system_1997}. We believe that in this regard, scaled graphs may provide additional flexibility in the analysis that traditional IQC methods cannot offer.

\section{Example}
\label{sec:lure_example}
In this section, we illustrate the effectiveness of the weighted scaled graph and its computation for Lur'e systems through an example.
\subsection{System Parameters}
\begin{figure}[b]
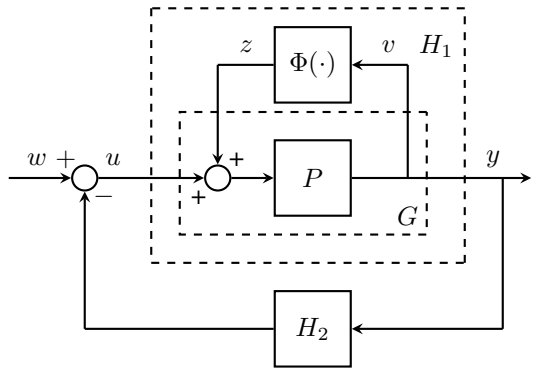

	\centering
	\ctikzfig{lure_example}
	\caption{Feedback interconnection considered in the example.}
	\label{fig:Lure_example}
\end{figure}

Consider the MIMO Lur'e system in Figure~\ref{fig:Lure_example}, where $P$ is a $2\times2$ LTI plant represented by the transfer function matrix
\begin{align} 
P(s) = \begin{bmatrix} P_{11}(s)&P_{12}(s)\\ P_{21}(s) & P_{22}(s) \end{bmatrix} ,
\end{align} 
with elements 
\begin{align*}
    P_{11}(s) = \frac{0.1}{s+1}, \; P_{12}(s) = \frac{1}{s^3+5s^2+2s+1}, \; P_{21}(s) = \frac{P_{12}(s)}{10},\; P_{22}(s) =\frac{0.2}{s+5}.
\end{align*}

For this example we have dimensions $p=q=2$, i.e., $\Phi$ is a $2\times 2$ system, and the system $G$ consistent with Figure~\ref{fig:final_NL_loop_transform} can be represented by 
\begin{align} \label{eq:H_lure_example}
\begin{bmatrix}
    v\\y
\end{bmatrix}= G\begin{bmatrix}
    z\\u
\end{bmatrix} = \begin{bmatrix} P_{11}&P_{12}&P_{11}&P_{12}\\
                    P_{21}&P_{22}&P_{21}&P_{22}\\
                    P_{11}&P_{12}&P_{11}&P_{12}\\
                    P_{21}&P_{22}&P_{21}&P_{22}\end{bmatrix}\begin{bmatrix}
    z\\u
\end{bmatrix}.
\end{align} 
Let $(A,B,C,D)$ denote a minimal state space representation of $G$ such that $G(s) = C(sI-A)^{-1}B+D$, where $B, C$ and $D$ are partitioned as
\begin{align*} 
B = \left[\!\begin{array}{c|c}
    B_z & B_u    
\end{array} \!\right], C = \left[\begin{array}{c}
    C_v \\ \hline C_y    
\end{array}\right], D = \left[ \begin{array}{c|c}
   D_{zv}  & D_uv \\ \hline
    D_{zy} & D_{uy} 
\end{array}\right].
\end{align*} 
We assume that  $\Phi$ satisfies the IQC in (\ref{eq:lure_IQC}) with 
$\Theta = \text{diag}( 1 , -0.1)$, i.e., $\Phi$ has an $\mathcal{L}_2$-gain of $\sqrt{0.1}$. Furthermore, we assume that $H_2 \! \in \! \mathcal{C}(\mathcal{D})$ where $\mathcal{D} \!= \! \left\{X \! \in \! \mathbb{S}^2 \!\mid \!X \!= \! \textup{diag}(x_1,x_2)\right\}$. 

\subsection{Numerical Results}
\begin{figure}[t]
    \centering
    \includegraphics[width=0.6\linewidth]{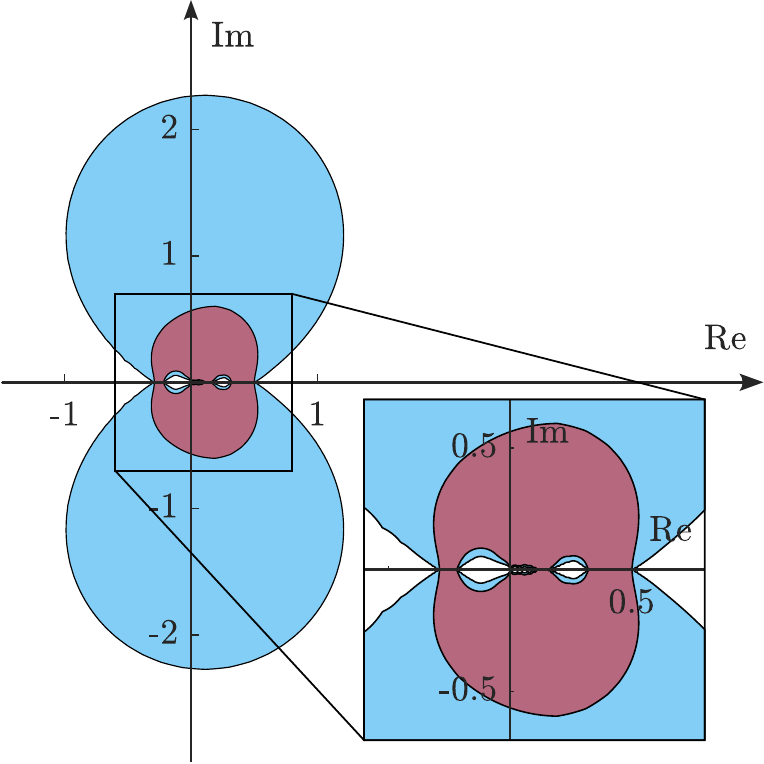}
    \caption{Over-approximation of $\textup{SG}(H_1)$ in blue, and over-approximation of $\textup{SG}^{W}(H_1)$, with $W = \text{diag}(1,16.3829)$ in purple.}
    \label{fig:plant_comparison}
\end{figure}

Using the approximation method of Theorem \ref{th:Lure_LMI} with $W=I$ (i.e., no scaling), we get (an over-approximation of) $\textup{SG}(H_1)$ indicated by the blue region shown in Figure~\ref{fig:plant_comparison}. Next, using the binary search, we find $W = \text{diag}(1,16.3829)$ for which the resulting (over-approximation of) $\textup{SG}^W(H_1)$ is indicated by the purple region in Figure~\ref{fig:plant_comparison}. For both the over-approximation of $\textup{SG}(H_1)$ and $\textup{SG}^W(H_1)$ the same values for $\lambda$ are used, and the same number of LMIs are solved, providing a fair comparison between approximations. Clearly, the region covered by $\textup{SG}^W(H_1)$ is significantly smaller than that covered by $\textup{SG}(H_1)$.

\begin{figure}[t]
\centering
\begin{subfigure}[t]{0.8\linewidth}
\centering
    \includegraphics[width=0.7\linewidth]{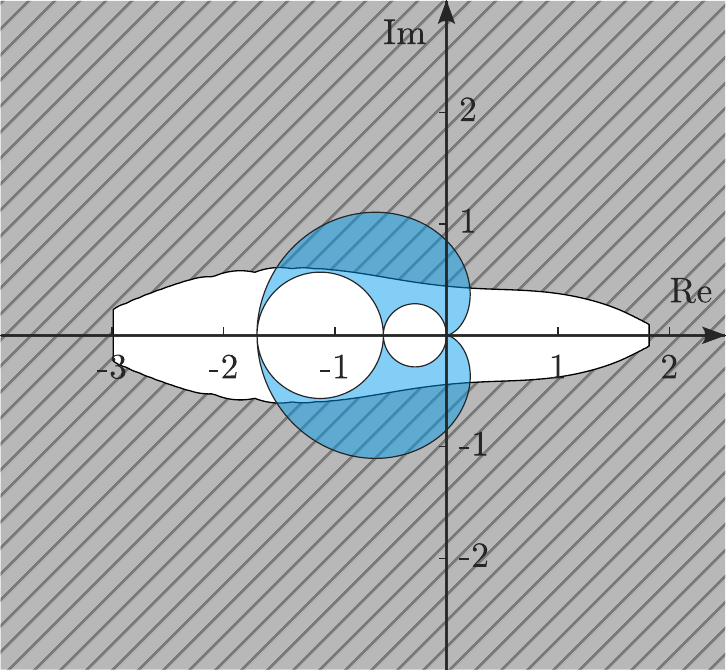}
    \caption{Over-approximation of $\text{SG}^{\dagger}(H_1)$ in gray, and $\text{SG}(-H_2)$ in blue. The regions overlap, and stability cannot be concluded directly using Theorem~\ref{th:SG}.}
    \label{fig:LTI_example_old}
\end{subfigure} \\
\begin{subfigure}[t]{0.8\linewidth}
\centering
    \includegraphics[width=0.7\linewidth]{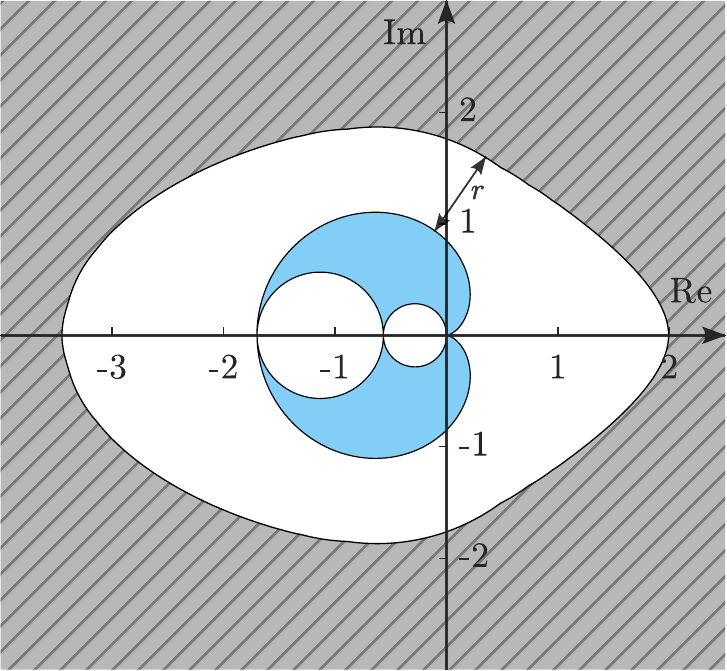}
    \caption{Over-approximation of $\text{SG}^{W\dagger}(H_1)$ with $W = \textup{diag}(1,16.3829)$ in gray, $\text{SG}(-H_2)$ in blue  and $r$ the shortest distance between $\text{SG}^{W\dagger}(H_1)$ and $\text{SG}(-H_2)$. The regions do not overlap and stability can be concluded using Theorem~\ref{th:SGW}.}
  \label{fig:LTI_example_new}
\end{subfigure}
\caption{Comparison between the use of scaled graphs (top) and weighted scaled graphs (bottom) for stability analysis of the Lur'e system in Figure~\ref{fig:Lure_example}.}
\label{fig:rms_comparison}
\end{figure}

To demonstrate the benefits of considering $\textup{SG}^W(H_1)$ over $\textup{SG}(H_1)$ in terms of reduced conservatism, we consider the inverse graphs $\textup{SG}(H_1)^\dagger$ and $\textup{SG}^{W\dagger}(H_1)$. These inverse graphs are depicted in Figure~\ref{fig:LTI_example_old} and Figure~\ref{fig:LTI_example_new}, respectively, by the gray-shaded regions. It is clear that $\textup{SG}^W(H_1)$ guarantees feedback stability for a much larger class of systems $H_2$ (e.g., with larger $\mathcal{L}_2$-gain) than concluded via $\textup{SG}(H_1)$, as the region covered by $\textup{SG}^{W\dagger}(H_1)$ is smaller in Figure~\ref{fig:LTI_example_new}) than the region covered by $\textup{SG}^{\dagger}(H_1)$ in Figure~\ref{fig:LTI_example_old}. 

To further illustrate this point, take as an example the diagonal LTI system $H_2$ represented by 
\begin{align} \label{eq:Lure_example_K}
H_2(s) = \begin{bmatrix} \frac{1.7}{s^2+2s+1}&0\\0&\frac{1.7}{s^2+3s+3} \end{bmatrix}
\end{align} 
for which the exact $\textup{SG}(H_2)$ is constructed via \cite[Theorem~1, Condition ii)]{pates_scaled_2021}, which for this specific case can be done since $H_2(s)H_2(s)^* = H_2(s)^*H_2(s)$, $s \in \mathbb{C}$. Clearly, using $\textup{SG}^\dagger(H_1)$ stability cannot be concluded, whereas through $\textup{SG}^{W\dagger}(H_1)$, which can be applied since $H_2 \in \mathcal{C}(\mathcal{D})$, stability is guaranteed. Furthermore, the smallest distance between $\textup{SG}^{W\dagger}(H_1)$ and $\textup{SG}^{W}(-H_2)$, indicated in Figure~\ref{fig:LTI_example_new} by $r$, can be interpreted as a robustness margin. Note that such margin does not easily follow from the LMIs. In conclusion, weighted scaled graphs  significantly improve  analysis of MIMO nonlinear feedback systems.

\section{Conclusions}\label{sec:conclusions}
In this paper, we exploit the use of multipliers in scaled graph analysis of MIMO systems. In particular, we use weighted inner products to arrive at a weighted version of the scaled graph. This weighted scaled graph may significantly reduce conservatism in the analysis of MIMO feedback systems having structure in their loop components. We present a method for computing the weighted scaled graphs for the specific class of Lur'e systems, and demonstrated the effectiveness in an example. Possible directions for future work include the extension to dynamic multipliers, as well as the extension of our computational methods to broader classes of nonlinear systems.

\section*{Appendix}
\appendix
\section{Proof of Theorem~\ref{th:Lure_LMI} }\label{app:A}
Take $\xi^\top = \begin{bmatrix} x^\top&\!\!\!\!\!\!z^\top &\!\!\!\!\!\!u^\top \end{bmatrix}$ and note that
\begin{align} 
\begin{bmatrix} z \\w \end{bmatrix} = \Gamma_w \xi , \quad \text{ and }\quad  \begin{bmatrix} y\\u \end{bmatrix} = \Gamma_y\xi,
\end{align} 
with $\Gamma_w$ and $\Gamma_y$ in \eqref{eq:Gamma_matrices}. Consider the transformation $\bar{u} = Xu$ and $\bar{y} = Xy$ such that we can write
\begin{equation}
       \begin{bmatrix} \bar{y}\\\bar{u} \end{bmatrix} = \begin{bmatrix} X&0\\0&X \end{bmatrix}   \Gamma_y\xi.
\end{equation}
Multiplying the matrix inequality in (\ref{eq:Lure_LMI}) from the left with $\xi^\top$ and from the right with $\xi$ results in 
\begin{align} \label{eq:storage_func}
\dot{S}(x) + \xi^\top \left( \Gamma_y^\top\Pi\otimes W\Gamma_y + \sigma\Gamma_w\Theta\otimes I_p \Gamma_w \right)\xi \leq 0,
\end{align} 
with $S(x) = x^\top Px$. Integrating (\ref{eq:storage_func}) from $t=0$ to $t=\infty$ yields
\begin{equation}
{S}(x(\infty))-S(x(0)) + \int_0^\infty\xi^\top  \Gamma_y^\top\Pi\otimes W\Gamma_y \xi \;dt \leq -\int_0^\infty\sigma \xi^\top\Gamma_w\Theta\otimes I_p \Gamma_w \xi \;dt\leq 0.
\end{equation}  
By assumption we have $x(0) = 0$ leading to $S(x(0))=0$. Moreover, since we assume that $H$ is stable, i.e., $H:\mathcal{L}_2^n\rightarrow\mathcal{L}_2^n$ ans solutions $x$ are absolutely continuous, we get for all $u\in \mathcal{L}_2^q$ that $\lim_{t\rightarrow\infty}S(x(t))=0$. Together with the assumption that $\Phi$ satisfies the IQC in \eqref{eq:lure_IQC}, this leads to the implication that
\begin{align} 
\int_0^\infty\xi^\top  \Gamma_y^\top\Pi\otimes W\Gamma_y \xi \;dt \leq 0.
\end{align} 
Using $W=X^\top X$ this expression can be written as
\begin{align} 
\int_0^\infty\xi^\top  \Gamma_y^\top \begin{bmatrix} X &0 \\ 0 & X \end{bmatrix}^\top  \Pi\otimes I_n \begin{bmatrix} X&0\\0&X \end{bmatrix}  \Gamma_y \xi \;dt \leq 0,
\end{align} 
or, equivalently,
\begin{align}\label{eq:proof_exp_2iqc}
\int_0^\infty \begin{bmatrix} \bar{y}\\ \bar{u} \end{bmatrix}^\top \Pi\otimes I_{n} \begin{bmatrix} \bar{y}\\ \bar{u} \end{bmatrix} \;dt \leq 0.   
\end{align} 
Note that \eqref{eq:proof_exp_2iqc} is an IQC expressed in terms of the input and output of the transformed plant $\bar{H}=XHX^{-1}$. Using \cite[Lemma 1]{van_den_eijnden_scaled_2024} we find that \eqref{eq:proof_exp_2iqc} implies 
\begin{align} 
\textup{SG}(\bar{H}) \subseteq \left\{z\in\mathbb{C}\;\left| \begin{bmatrix} z \\ 1 \end{bmatrix}^* \Pi \begin{bmatrix} z \\ 1 \end{bmatrix} \leq 0     \right.\right\}.
\end{align} 
This completes the proof. \hfill $\square$

\bibliographystyle{ieeetr}   
{\footnotesize
\bibliography{Refs_reduced}
}

\end{document}